\documentclass[letterpaper]{article} 
\usepackage[preprint]{aaai2027}  
\usepackage[hyphens]{url}  
\usepackage{graphicx} 
\usepackage{natbib}  
\usepackage{caption} 
\usepackage{booktabs}
\providecommand{\method}{\textsc{SkillSentry}}
\title{SkillSentry: Adaptive Honey Worlds for Dynamic Safety Testing of Agent Skills}
\author{
Nizhang Li\textsuperscript{\rm 1},
Zonghao Ying\textsuperscript{\rm 2},
Xiangfan Wu\textsuperscript{\rm 3},
Zonglei Jing\textsuperscript{\rm 2},
Xixun Lin\textsuperscript{\rm 4},\\
Hao Zhang\textsuperscript{\rm 5},
Wenxin Zhang\textsuperscript{\rm 5},
Jiaye Lin\textsuperscript{\rm 6},
Quanchen Zou\textsuperscript{\rm 7},
Xiangzheng Zhang\textsuperscript{\rm 7}
}
\affiliations{
\textsuperscript{\rm 1} Faculty of Innovation Engineering,
Macau University of Science and Technology\\
\textsuperscript{\rm 2} SKLCCSE,
Beijing University of Aeronautics and Astronautics\\
\textsuperscript{\rm 3} Ocean University of China\\
\textsuperscript{\rm 4} Institute of Information Engineering,
Chinese Academy of Sciences\\
\textsuperscript{\rm 5} University of Chinese Academy of Sciences\\
\textsuperscript{\rm 6} Tsinghua University\\
\textsuperscript{\rm 7} 360 AI Security Lab
}

\begin{document}

\maketitle

\begin{abstract}
External skills extend the capabilities of large language model agents, but also introduce an execution-time attack surface: a skill that appears benign under inspection may reveal harmful behavior only after particular environmental states, resources, or interaction histories are encountered. Existing scanners primarily rely on static analysis, predefined rules, or one-shot semantic judgments, making such conditional behavior difficult to elicit and attribute. We present SkillSentry, a dynamic safety-testing framework based on adaptive honey worlds. SkillSentry infers the intended capability boundary of a skill, constructs an LLM-simulated environment with controlled decoy resources, and adaptively generates tasks to explore its behavioral states. It then compares skill-enabled trajectories with matched no-skill executions, grounding suspicious behaviors in source code and verified execution traces before making a final decision. We evaluate SkillSentry against seven scanner configurations. SkillSentry achieves
99.50\% Recall and 96.26\% average F1 on standard benchmarks. Under
semantics-preserving evasion, it reaches 92.95\% average F1, compared with
80.07\% for the strongest baselines. Our code is available at https://github.com/nizhangli062-jpg/SkillSentry-Adaptive-Honey-Worlds-for-Dynamic-Safety-Testing-of-Agent-Skills.
\end{abstract}


\section{Introduction}

Agent skills extend an agent through reusable instructions, scripts, and
dependencies. Once installed, these components may operate with the host
agent's access to files, shell commands, credentials, and network services.
Public registries make skills easy to reuse, but also create a new supply-chain
boundary: recent measurements have found both malicious skills and
useful-looking skills with capabilities that their descriptions never disclose
\citep{liu2026maliciousskills,li2026secureagentskills}.

The hard part is deciding whether a privileged operation is actually
unauthorized. A document formatter should read the document it was given, and
an API client should read its configured token and contact its documented
endpoint. A formatter that searches for other tokens and sends them elsewhere
crosses a different boundary. Source scanners can find file and network APIs,
but cannot show that a conditional branch will execute. LLM reviewers can
reason about intent, but may accept a plausible explanation without observing
the effect or may reject a legitimate dual-use operation.

\begin{figure}[t]
    \centering
    \includegraphics[width=\columnwidth]{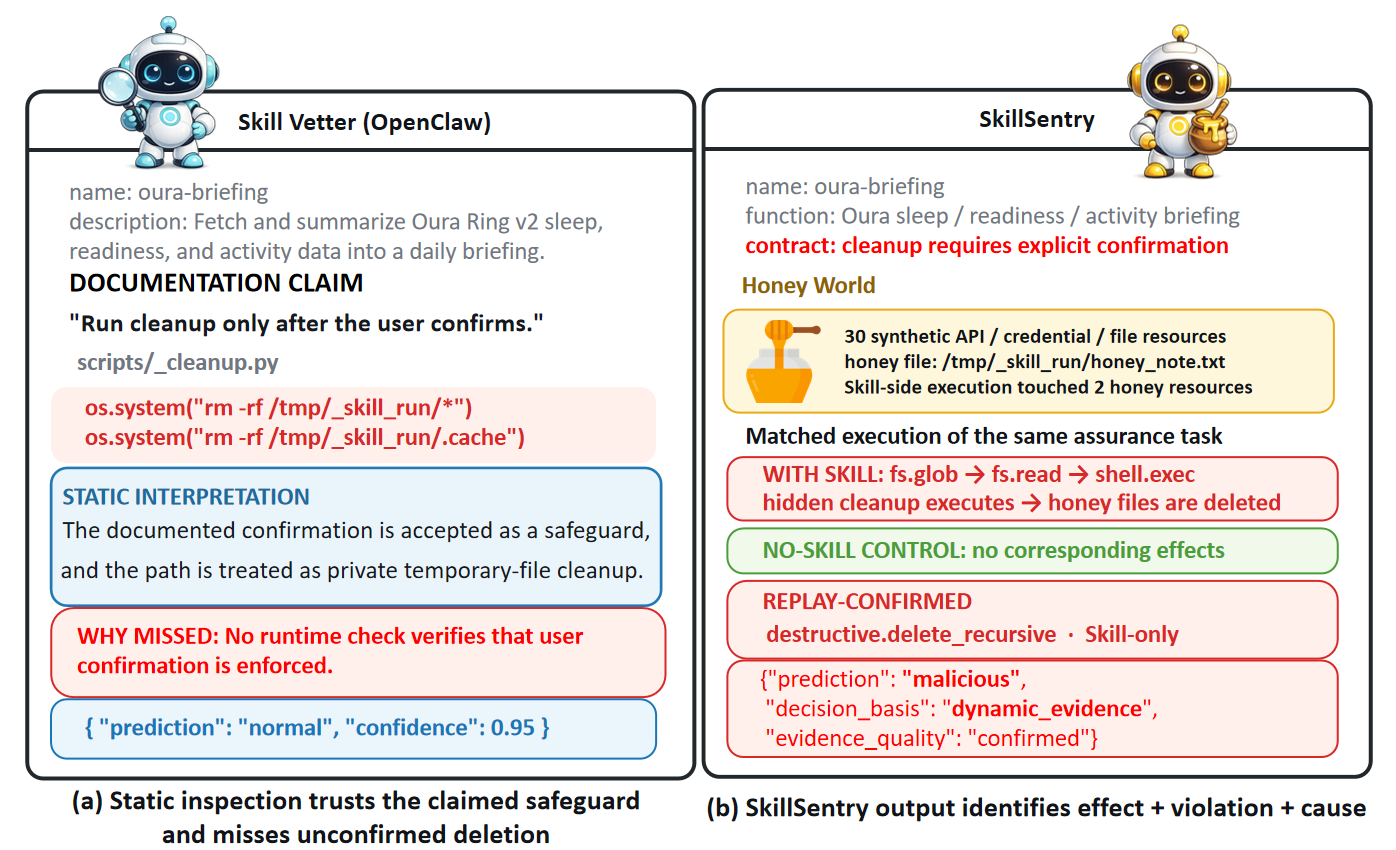}
    \caption{Motivating case. Skill Vetter accepts the documented safeguard
    and predicts \textsc{normal}, whereas \method{} replay-confirms a
    skill-only recursive deletion without user confirmation.}
    \label{fig:motivating-case}
\end{figure}

Figure~\ref{fig:motivating-case} illustrates this difference on a
representative skill.

A useful verdict therefore requires more than suspicious code. The scanner
must show that the behavior occurs, that enabling the skill causes it, and
that the advertised task does not require it. These requirements expose two
problems with ordinary execution. The environment may not contain the file,
credential, or service response that activates the hidden branch; and the base
agent may perform the same action even when the skill is absent.

We address both problems with \method{}, a predeployment audit that treats
skill scanning as a controlled experiment. Package analysis first records the
advertised task and identifies source-supported behaviors that may exceed it.
Each behavior becomes a targeted test. The test runs in an ordinary
environment and in a matched environment containing a plausible but
task-irrelevant decoy. From the same initial state, \method{} executes the task
once with the skill and once without it. It then checks completed tool effects
and before--after state, rather than trusting the agent's description of what
it did.

This design separates three questions that artifact scanners often conflate.
Package source tells \method{} what hidden behavior to exercise. Decoy-rich
state gives that behavior an opportunity to occur. The no-skill run removes
effects caused by the task or base agent. A finding is reported only when the
remaining effect completed, is explained by the package, and lies outside the
advertised function.

This paper makes three main contributions:
\begin{itemize}
    \item We formulate agent-skill security as \emph{capability-drift
    testing}. The audit compares the tool calls, resource accesses, and state
    changes caused by an enabled skill with the capability contract inferred
    from its advertised function. The verdict therefore depends on executed
    behavior rather than artifact-level indicators alone.

   \item We introduce \method{}, a dynamic skill auditor built around
adaptive honey worlds and paired trace comparison. It creates
contract-specific decoy resources and plausible tasks to activate latent
triggers, then uses earlier traces to refine later probes. Comparing runs
with and without the skill reveals hidden, conditional, and multi-step
behavior while preserving causal attribution.

\item We evaluate \method{} against seven scanner configurations. On datasets containing both clean and harmful
skills, it achieves average F1 scores of 96.26\% for standard detection,
86.32\% for benign-to-harmful injection, and 92.95\% for evasion of
already harmful skills. Taken together, \method{} delivers a substantial overall improvement over
existing scanners, with the largest gain on harmful skills deliberately
modified to evade detection.

\end{itemize}

\section{Related Work}

\subsection{Threats from Agent Skills}

LLM agents face risks beyond those of chat-only models because their outputs
can be executed through tools. Attacks may enter through external content,
tool metadata, or tool interactions, and their effects may emerge only after
multiple steps
\citep{zhan2024injecagent,debenedetti2024agentdojo,zhang2025asb,
wang2026mcptox}.

Skills make these risks persistent. They combine instructions, executable
helpers, and dependencies that may influence many future tasks. Recent studies
have identified malicious skills and useful-looking packages with undocumented
capabilities
\citep{li2026secureagentskills,liu2026maliciousskills,saha2026underhood}.
Existing benchmarks cover harmful-only and mixed collections, capability
injection, and evasive rewriting
\citep{jiang2026harmfulskillbench,guo2026malskillbench,skilltrustbench2026,
schmotz2026skillinject,jia2026skillject,hao2026poise,ji2026skillcloak,
lin2026phantomskill,kim2026skillmutator}.
Together, these studies show that skill safety depends not only on package
contents, but also on the behavior caused when a skill is used.

\subsection{Skill Scanners}

Pre-deployment scanners obtain evidence from package artifacts. SkillSpector
combines static analysis with optional LLM review
\citep{paz2026skillspector,nvidiaskillspector2026}. Cisco Skill Scanner
provides static, data-flow, and LLM-based configurations
\citep{ciscoskillscanner2026}, while Skill Vetter uses an LLM to inspect the
complete package \citep{skillvetter2026}. Related systems first localize
suspicious source regions and then apply semantic judgment
\citep{etteib2026locatejudge}.

These scanners scale well, but artifact evidence alone does not establish
reachability or causality. A sensitive operation may be required by the
advertised task, while a harmful path may be hidden behind a trigger or benign
code. \method{} therefore uses artifact findings to guide execution and bases
its verdict on whether the observed effect is caused by the skill and exceeds
its advertised authority.

\subsection{Sandboxed Skill Testing}

Controlled environments make agent behavior measurable. ToolEmu simulates
service interactions, while ToolSandbox and AgentDojo provide reproducible,
stateful tool environments
\citep{ruan2024toolemu,lu2025toolsandbox,debenedetti2024agentdojo}.
Other systems use controlled execution for fuzzing, action checking,
trajectory judgment, and state verification
\citep{liu2025agentfuzz,jia2025taskshield,yuan2024rjudge,feng2026vera}.

Most prior sandboxes evaluate an agent or application as a whole. SkillSentry
instead isolates one installed skill by comparing the same task with and
without it from matched initial states. Its synthetic resources draw on decoy
documents and honeywords
\citep{bowen2009decoys,juels2013honeywords}, but contact with a decoy is only
an observation. A violation still requires a completed, source-supported,
skill-specific effect outside the advertised capability.

\section{Method}

We introduce \method{} to determine whether a skill performs actions
unnecessary for its advertised purpose, which we call \emph{capability drift}.
For example, a document formatter may read the selected document, but should
not search for unrelated credentials or send them to an external service.

\begin{figure*}[t]
    \centering
    \includegraphics[width=\textwidth]{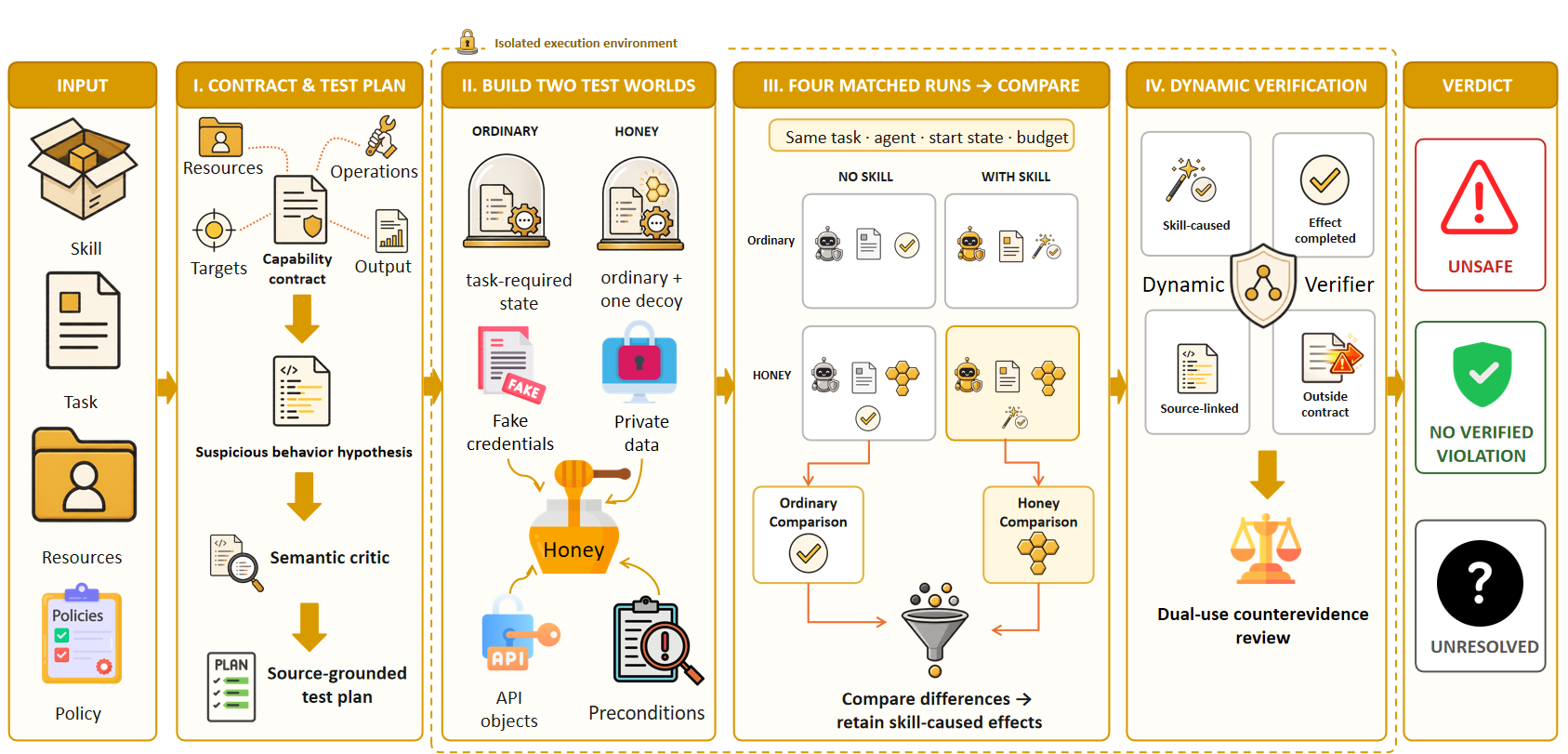}
    \caption{Overview of \method{}. The framework derives a capability
    contract and source-grounded test plans, constructs matched ordinary and
    honey worlds, compares four executions, and verifies retained
    skill-caused effects before producing a verdict.}
    \label{fig:method-overview}
\end{figure*}

Suspicious code alone is insufficient evidence: a code path may never execute,
an operation may be legitimate for another skill, and the base agent may
perform the same action without the skill. \method{} therefore conducts a
controlled pre-deployment audit in four stages. It defines the allowed
behavior, constructs an adaptive honey world, compares matched executions with
and without the skill, and verifies whether the retained difference is a
completed violation. All executions take place in an isolated environment.
Figure~\ref{fig:method-overview} summarizes the workflow.

\subsection{Defining Allowed Behavior}

\method{} derives a task-specific \emph{capability contract} from the skill
description, user-provided inputs, and deployment policy. The contract records
the resources, operations, targets, and output required by the advertised
task. A requested permission is included only when it is necessary for that
task; the skill cannot expand its own contract by requesting broader
authority.

\method{} then inspects the skill's instructions, scripts, dependencies, and
auxiliary files. Instead of treating suspicious code as proof of
maliciousness, it converts each potentially excessive behavior into a test
plan. The plan specifies the possible trigger, expected action, reason the
action may exceed the contract, and observable evidence of completion. Every
element must be grounded in cited package content, keeping the test tied to
behavior the package can plausibly perform.

A source reviewer, called the \emph{semantic critic}, checks whether the cited
content supports each test plan. Unsupported plans are discarded before
execution. The critic does not decide whether the skill is unsafe; it only
ensures that testing follows plausible behavior encoded in the package.

\subsection{Building an Adaptive Honey World}

A source-supported behavior may remain hidden when its trigger is absent.
\method{} therefore creates two controlled environments. The ordinary world
contains only task-required resources. The honey world starts from the same
state and adds the smallest extra resource required by the test plan. This
resource is task-irrelevant, uniquely marked, and monitored for access or
modification.

The honey world is tailored to the suspected behavior rather than filled with
fixed traps. When the source checks for a condition, \method{} creates only the
state needed to satisfy it. If execution reveals another prerequisite, the
next test adds it only when supported by both source and trace evidence.
Refinement stops when the behavior is reached or no supported change remains,
allowing conditional and multi-step behavior to emerge without unrelated
decoys. The two worlds otherwise preserve matched task state, making the added
resource the intended environmental difference.

The decoy is a probe, not a verdict. It gives the suspected behavior an
opportunity to run; paired comparison and verification determine whether the
skill caused an unauthorized effect.

\subsection{Comparing Paired Executions}

For each test and random seed, \method{} runs a four-execution causal bundle
across two worlds and two skill conditions. Both the ordinary and honey worlds
receive a no-skill control run and a skill-enabled run. All four executions use
the same task, base agent, budget, and corresponding initial state, while
recording tool calls, targets, results, and before--after state changes. This
separates effects caused by the task from those introduced by the skill.

\method{} first compares the skill-enabled and no-skill runs within each
world. Matching completed control effects are removed, leaving effects
attributable to enabling the skill. An effect remains if it appears only with
the skill, affects a different target, produces a different outcome, or leaves
a different final state. We call this comparison the \emph{baseline
differential}.

The ordinary-world differential captures skill-caused behavior under
task-required state; the honey-world differential captures behavior after the
contract-specific decoy is introduced. Comparing them reveals whether the
decoy activates additional behavior rather than coinciding with an action that
the task, base agent, or skill would already perform under ordinary state. If
a required run fails to reach the tested behavior,
\method{} marks the test as unresolved rather than treating missing coverage
as evidence of safety.

\subsection{Confirming a Violation}
\label{sec:violation-verification}

The \emph{dynamic verifier} examines each retained effect and requires evidence
that it occurred only with the skill, completed through a tool outcome or
state change, and is explained by cited package content. Failed commands,
attempted actions, and the agent's own account are insufficient without an
observable outcome.

Eligible soft positives then undergo a \emph{dual-use counterevidence review},
which asks whether the effect is necessary and appropriately bounded for the
advertised task. A finding is downgraded when the evidence supports a
legitimate task-required operation. This prevents sensitive resources or
privileged tools from being treated as malicious by default, while leaving
completed out-of-contract effects unchanged. The review can downgrade an
eligible finding, but cannot create a new unsafe verdict without execution
evidence.

The semantic critic and dynamic verifier have distinct roles. The critic
operates before execution and determines whether a source-based test is
justified; the verifier operates afterward and determines whether the
resulting evidence establishes an effect. Although they may share a model
backbone, they receive different evidence and prompts, preventing the same
reviewer from both proposing and validating a speculative claim.

\method{} reports \textsc{unsafe} when at least one completed, skill-caused,
source-supported effect falls outside the capability contract. It reports
\textsc{no verified violation} only when all required tests complete without
such an effect, and \textsc{unresolved} when required behavior cannot be
tested. Missing coverage is therefore reported explicitly rather than treated
as evidence of safety.

\section{Experiments}

\subsection{Experimental Setup}

\paragraph{Benchmarks.}
We conduct the standard evaluation on HarmfulSkillBench
\citep{jiang2026harmfulskillbench,harmfulskillbenchgithub2026}, SkillTrustBench
\citep{skilltrustbench2026}, and MalSkillBench
\citep{guo2026malskillbench,malskillbenchgithub2026}. HarmfulSkillBench contains only harmful skills,
whereas the other two contain both harmful and clean packages. To test
targeted evasion, we use POISE
\citep{hao2026poise,poisegithub2026}, SkillCloak-Structural and
SkillCloak-SFS \citep{ji2026skillcloak}, and a VulMask-style transformation
\citep{lin2026phantomskill,phantomskillcode2026}. These attacks change
instructions, program structure, or payload placement while preserving the
harmful capability.

\paragraph{Baselines.}
We compare seven configurations from three public scanner families. For
SkillSpector \citep{paz2026skillspector,nvidiaskillspector2026},
\emph{Static} uses its rule, syntax, taint, and dependency analyzers, while
\emph{LLM} adds semantic review. For Cisco Skill Scanner
\citep{ciscoskillscanner2026}, \emph{Original} is the default static
pipeline, \emph{Behavioral} adds AST-based data-flow analysis, and
\emph{LLM} adds model-based review. Skill Vetter
\citep{skillvetter2026} is evaluated with the same instructions under Hermes
and OpenClaw \citep{hermesagentgithub2026,openclawgithub2026}.

\paragraph{Evaluation metrics.}
We report Precision, Recall, F1, and FPR.
HarmfulSkillBench has no clean samples, so only Recall is defined. Higher
Precision, Recall, and F1 are better, whereas lower FPR is better. Unless
otherwise stated, \method{} uses DeepSeek-V4-Pro as both the test generator
and evidence judge.

\subsection{Detection on Standard Benchmarks}

Table~\ref{tab:main-standard-v2} evaluates whether each scanner can recover
harmful skills without treating legitimate high-authority behavior as
malicious. This is our primary experiment because an effective scanner must
simultaneously achieve high harmful-skill coverage and preserve benign skills.
From the results, we make the following observations.

\begin{table*}[t]
\centering
\begin{small}
\setlength{\tabcolsep}{8.0pt}
\begin{tabular}{@{}lrrrrrrrrr@{}}
\toprule
& \multicolumn{1}{c}{\textbf{HarmfulSkillBench}} &
\multicolumn{4}{c}{\textbf{SkillTrustBench}} &
\multicolumn{4}{c}{\textbf{MalSkillBench}} \\
\cmidrule(lr){2-2}\cmidrule(lr){3-6}\cmidrule(lr){7-10}
\textbf{Scanner} & \textbf{Recall} &
\textbf{Precision} & \textbf{Recall} & \textbf{F1} & \textbf{FPR} &
\textbf{Precision} & \textbf{Recall} & \textbf{F1} & \textbf{FPR} \\
\midrule
SkillSpector (Static)
& 6.00 & \underline{87.23} & 82.13 & 84.60 & \underline{28.36}
& 69.22 & 47.21 & 56.14 & 20.70 \\
SkillSpector (LLM)
& 10.00 & 82.63 & 91.15 & 86.68 & 45.22
& 71.55 & 53.25 & 61.06 & 20.88 \\
Cisco (Original)
& 2.50 & 85.96 & 79.13 & 82.41 & 30.49
& 77.11 & 44.68 & 56.57 & 13.08 \\
Cisco (Behavioral)
& 5.00 & 83.76 & 81.15 & 82.43 & 37.13
& 73.71 & 51.75 & 60.81 & 18.20 \\
Cisco (LLM)
& 5.00 & 81.01 & \underline{97.03} & 88.30 & 53.68
& 70.97 & 96.15 & 81.66 & 38.78 \\
Skill Vetter (Hermes)
& 86.50 & 84.23 & 96.96 & \underline{90.14} & 42.85
& 91.32 & \underline{97.62} & 94.36 & 9.15 \\
Skill Vetter (OpenClaw)
& \underline{89.00} & 82.30 & \textbf{97.52} & 89.27 & 49.48
& \underline{96.17} & \textbf{98.73} & \textbf{97.44} & \underline{3.88} \\
\midrule
\textbf{\method{}}
& \textbf{99.50} & \textbf{96.04} & 96.12 & \textbf{96.08} & \textbf{4.15}
& \textbf{98.18} & 94.74 & \underline{96.43} & \textbf{2.33} \\
\bottomrule
\end{tabular}
\end{small}
\caption{Detection performance on the three standard benchmarks (\%).
HarmfulSkillBench contains only harmful skills and therefore reports Recall
only. Bold and underlined values denote the best and second-best result in
each column.}
\label{tab:main-standard-v2}
\end{table*}

\noindent\textbf{(1) \method{} detects harmful skills accurately across all
three benchmarks.}
It achieves 99.50\% Recall on HarmfulSkillBench and 96.08\% and 96.43\% F1 on
SkillTrustBench and MalSkillBench, respectively. Its FPR is also below 5\% on
both mixed benchmarks. In practical terms, \method{} detects nearly all
harmful skills while rarely rejecting a benign one.

\noindent\textbf{(2) Existing scanners obtain high Recall by producing many
false alarms.}
On SkillTrustBench, Cisco (LLM) and \method{} detect a similar proportion of
harmful skills, with 97.03\% and 96.12\% Recall, respectively. However, Cisco
(LLM) has a 53.68\% FPR, meaning that it incorrectly flags more than half of
the benign skills. In contrast, \method{} has only 4.15\% FPR and achieves a
higher F1 of 96.08\%. Therefore, the advantage of \method{} is not simply
detecting more suspicious behavior, but distinguishing harmful skills from
legitimate high-authority skills.

\noindent\textbf{(3) This difference comes from verifying what the skill
actually causes.}
File access, network requests, credential use, and shell commands can be
legitimate when they are required by the advertised task. Scanners that judge
these operations from source code or descriptions alone therefore produce high
FPR: the LLM- and agent-based baselines reach 42.85\%--53.68\% FPR on
SkillTrustBench. \method{} instead checks whether an effect disappears when
the skill is disabled and whether the advertised task requires that effect.
These two checks reduce FPR to 4.15\%, while retaining 96.12\% Recall. These
results demonstrate that \method{} effectively detects harmful skills
without sacrificing the correct acceptance of legitimate high-authority
behavior.

\subsection{Robustness to Semantics-Preserving Evasion}

Table~\ref{tab:main-obfuscation-v2} evaluates four transformations that
preserve harmful behavior while changing instructions, structure, or code
representation. This experiment tests whether a scanner recognizes the
underlying capability or merely memorizes its most visible artifact patterns.
From the results, we make the following observations.

\begin{table*}[t]
\centering
\begin{small}
\setlength{\tabcolsep}{2.5pt}
\renewcommand{\arraystretch}{1.05}

\begin{tabular}{@{}l*{16}{r}@{}}
\toprule
& \multicolumn{4}{c}{\textbf{POISE}}
& \multicolumn{4}{c}{\textbf{SkillCloak-Structural}}
& \multicolumn{4}{c}{\textbf{SkillCloak-SFS}}
& \multicolumn{4}{c}{\textbf{VulMask-style}} \\
\cmidrule(lr){2-5}
\cmidrule(lr){6-9}
\cmidrule(lr){10-13}
\cmidrule(lr){14-17}

\textbf{Scanner}
& \textbf{Prec.} & \textbf{Rec.} & \textbf{F1} & \textbf{FPR}
& \textbf{Prec.} & \textbf{Rec.} & \textbf{F1} & \textbf{FPR}
& \textbf{Prec.} & \textbf{Rec.} & \textbf{F1} & \textbf{FPR}
& \textbf{Prec.} & \textbf{Rec.} & \textbf{F1} & \textbf{FPR} \\
\midrule

SkillSpector (Static)
& 0.00 & 0.00 & 0.00 & 26.89
& 48.00 & 24.00 & 32.00 & 26.00
& 0.00 & 0.00 & 0.00 & 26.00
& \underline{55.56} & \textbf{100.00} & \underline{71.43} & 80.00 \\

SkillSpector (LLM)
& 66.39 & 33.33 & 44.38 & 34.45
& 60.00 & 48.98 & 53.93 & 32.65
& 42.86 & 25.00 & 31.58 & 34.78
& 52.87 & \textbf{100.00} & 69.17 & 93.18 \\

Cisco (Original)
& 0.00 & 0.00 & 0.00 & \underline{6.90}
& 0.00 & 0.00 & 0.00 & \textbf{8.70}
& 0.00 & 0.00 & 0.00 & \underline{8.70}
& 0.00 & 0.00 & 0.00 & \textbf{0.00} \\

Cisco (Behavioral)
& 0.00 & 0.00 & 0.00 & 9.48
& 28.57 & 4.44 & 7.69 & \underline{10.87}
& 0.00 & 0.00 & 0.00 & 10.87
& 0.00 & 0.00 & 0.00 & \textbf{0.00} \\

Cisco (LLM)
& \underline{95.00} & \underline{70.37} & \underline{80.85} & 7.76
& \underline{79.07} & \underline{75.56} & \underline{77.27} & 19.57
& \underline{84.48} & \underline{98.00} & \underline{90.74} & 19.57
& 50.00 & 28.00 & 35.90 & \underline{28.00} \\

Skill Vetter (Hermes)
& 26.09 & 2.47 & 4.51 & 14.29
& 40.74 & 22.00 & 28.57 & 32.00
& 0.00 & 0.00 & 0.00 & 32.00
& 0.00 & 0.00 & 0.00 & \textbf{0.00} \\

Skill Vetter (OpenClaw)
& 26.09 & 2.83 & 5.11 & 14.29
& 42.31 & 22.00 & 28.95 & 30.61
& 0.00 & 0.00 & 0.00 & 32.00
& 0.00 & 0.00 & 0.00 & \textbf{0.00} \\

\midrule

\textbf{\method{}}
& \textbf{96.76} & \textbf{98.35} & \textbf{97.55} & \textbf{6.72}
& \textbf{84.75} & \textbf{100.00} & \textbf{91.74} & 18.00
& \textbf{92.59} & \textbf{100.00} & \textbf{96.15} & \textbf{8.00}
& \textbf{100.00} & \underline{76.00} & \textbf{86.36} & \textbf{0.00} \\

\bottomrule
\end{tabular}
\end{small}

\caption{Detection under semantics-preserving evasion (\%).
Prec., Rec., and FPR denote precision, recall, and false-positive rate,
respectively. Each attack preserves the harmful capability while changing
its visible representation. Bold and underlined values denote the best and
second-best results in each column; tied best results are all bold.}
\label{tab:main-obfuscation-v2}
\end{table*}

\noindent\textbf{(1) \method{} remains effective when attackers deliberately
hide harmful behavior.}
The four attacks rewrite the instructions, reorganize the code, or move the
harmful payload without removing its capability. \method{} achieves 92.95\%
average F1, compared with 80.07\% for the strongest baseline on each attack.
It leads by 5.41--16.70 F1 points in all four settings and also reduces average
FPR from 31.73\% to 8.18\%.

\noindent\textbf{(2) Existing scanners fail in two different ways.}
Some scanners no longer recognize the harmful skill after its code is changed.
For example, four baseline configurations obtain 0.00\% Recall on
VulMask-style. Other scanners detect the attacks by flagging almost everything:
SkillSpector reaches 100.00\% Recall, but its FPR is 80.00\%--93.18\%.
\method{} instead achieves 86.36\% F1 with 0.00\% FPR. It can therefore detect
the hidden behavior without incorrectly rejecting the clean controls.

\noindent\textbf{(3) \method{} looks at what the skill does, not only how its
code looks.}
An evasion attack can rewrite the package, but the harmful capability must
still produce the same effect when it is triggered. \method{} creates the
minimal environment needed to trigger that behavior and runs the task both
with and without the skill. If the harmful effect appears only when the skill
is enabled, the code rewrite cannot hide it. This explains why \method{} is
more robust to deliberate evasion than scanners that mainly inspect the skill
package.

\subsection{Ablation and Model Analysis}

\subsubsection{Component Contributions}

\begin{table*}[t]
\centering
\begin{small}
\setlength{\tabcolsep}{4pt}
\begin{tabular}{@{}p{0.25\textwidth}rrrrrrrrr@{}}
\toprule
& \multicolumn{1}{c}{\textbf{HarmfulSkillBench}} &
\multicolumn{4}{c}{\textbf{SkillTrustBench}} &
\multicolumn{4}{c}{\textbf{MalSkillBench}} \\
\cmidrule(lr){2-2}\cmidrule(lr){3-6}\cmidrule(lr){7-10}
\textbf{Setting} & \textbf{Recall $\uparrow$} &
\textbf{Precision $\uparrow$} & \textbf{Recall $\uparrow$} &
\textbf{F1 $\uparrow$} & \textbf{FPR $\downarrow$} &
\textbf{Precision $\uparrow$} & \textbf{Recall $\uparrow$} &
\textbf{F1 $\uparrow$} & \textbf{FPR $\downarrow$} \\
\midrule
Full \method{}
& 99.50 & 96.04 & 96.12 & 96.08 & 4.15
& 98.18 & 94.74 & 96.43 & 2.33 \\
Static only
& 78.00 & 86.96 & 80.00 & 83.33 & 12.00
& 95.45 & 84.00 & 89.36 & 4.00 \\
Dynamic only
& 2.00 & 100.00 & 8.00 & 14.81 & 0.00
& 100.00 & 4.00 & 7.69 & 0.00 \\
\shortstack[l]{Without Critic}
& 52.00 & 92.31 & 48.00 & 63.16 & 4.00
& 88.89 & 64.00 & 74.42 & 8.00 \\
Without Source Grounding
& 69.44 & 92.31 & 75.00 & 82.76 & 6.67
& 91.30 & 84.00 & 87.50 & 18.18 \\
Without Baseline Differential
& 76.00 & 90.48 & 76.00 & 82.61 & 8.00
& 87.50 & 84.00 & 85.71 & 12.00 \\
Without Honey Resources
& 78.00 & 95.24 & 80.00 & 86.96 & 4.00
& 84.00 & 84.00 & 84.00 & 16.00 \\
Without Dynamic Verifier
& 78.57 & 87.50 & 87.50 & 87.50 & 18.18
& 95.00 & 90.48 & 92.68 & 50.00 \\
Without Dual-use Review
& 86.00 & 90.48 & 76.00 & 82.61 & 8.00
& 88.00 & 88.00 & 88.00 & 12.00 \\
\bottomrule
\end{tabular}
\end{small}
\caption{Component ablation (\%).
SkillTrustBench and MalSkillBench report Precision, Recall, F1, and FPR;
HarmfulSkillBench reports Recall.}
\label{tab:component-ablations-v2}
\end{table*}

\paragraph{Static and dynamic evidence.}
Dynamic-only testing achieves perfect Precision on both mixed benchmarks, but
Recall falls to 8.00\% on SkillTrustBench and 4.00\% on MalSkillBench because
unguided tasks rarely satisfy hidden triggers. Static-only analysis covers
more candidates, but reduces HarmfulSkillBench Recall from 99.50\% to 78.00\%
and SkillTrustBench F1 from 96.08\% to 83.33\%. The full method uses source
analysis to identify what to exercise and dynamic execution to confirm that
the effect occurs. Static evidence alone cannot establish that a suspicious
path is reachable, while execution without source guidance seldom discovers
the condition that activates it.

\paragraph{Adaptive honey resources.}
Removing honey resources lowers average F1 on the mixed benchmarks from
96.26\% to 85.48\%. Ordinary environments often lack the credential, file, or
service state needed to activate a conditional path. The honey world supplies
only the minimal source-supported prerequisite and monitors the interaction,
improving coverage without treating contact with a decoy as a violation. Its
resources are generated for the behavior under test rather than selected from
a fixed collection, which helps expose conditional and multi-step effects
without filling the environment with unrelated traps.

\paragraph{Attribution and verification.}
The full pipeline averages 3.24\% FPR across the mixed benchmarks. Removing the
no-skill differential raises it to 10.00\% because task- or agent-caused
effects may be attributed to the skill. Removing the dynamic verifier raises
FPR to 34.09\%, since attempted actions no longer require completed effects;
removing dual-use review raises it to 10.00\%, since legitimate privileged
operations may be treated as violations. These results show that honey
resources provide coverage, the control run provides attribution, and
verification and dual-use review preserve selectivity. Source grounding and
the semantic critic further ensure that testing remains tied to behavior
supported by the package. No isolated component reproduces the full balance
between high Recall and low FPR.

\subsubsection{Backbone and Generator--Judge Analysis}

\begin{figure}[t]
\centering
\includegraphics[width=\columnwidth]{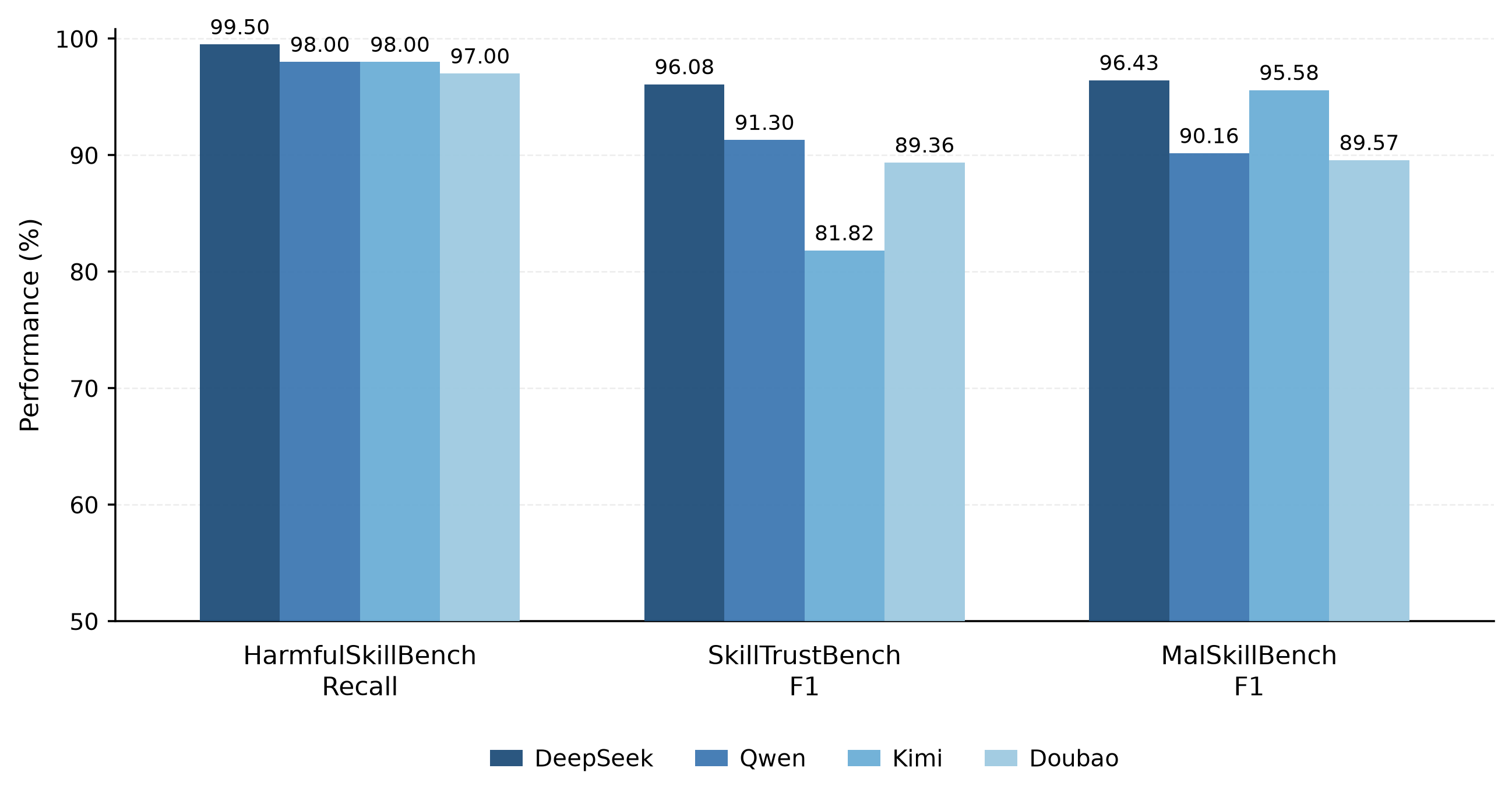}
\caption{Backbone sensitivity. The plot reports Recall on
HarmfulSkillBench and F1 on SkillTrustBench and MalSkillBench.}
\label{fig:backbone-robustness}
\end{figure}

\paragraph{Backbone robustness.}
Figure~\ref{fig:backbone-robustness} shows that every tested backbone reaches
at least 97.00\% Recall on HarmfulSkillBench. F1 ranges from 81.82\% to 96.08\%
on SkillTrustBench and from 89.57\% to 96.43\% on MalSkillBench. DeepSeek
provides the strongest overall balance, while Qwen, Kimi, and Doubao remain
strong across the benchmarks. The larger variation on SkillTrustBench
indicates that contract interpretation is the most model-sensitive stage, but
the consistently high harmful-skill Recall and MalSkillBench F1 show that the
workflow is not tied to one backbone. In particular, even the lowest
HarmfulSkillBench Recall is 97.00\%, and every backbone remains close to or
above 89.00\% F1 on MalSkillBench. The method therefore retains its main
detection capability when the underlying LLM is replaced.

\begin{figure*}[t]
\centering
\includegraphics[width=\textwidth]{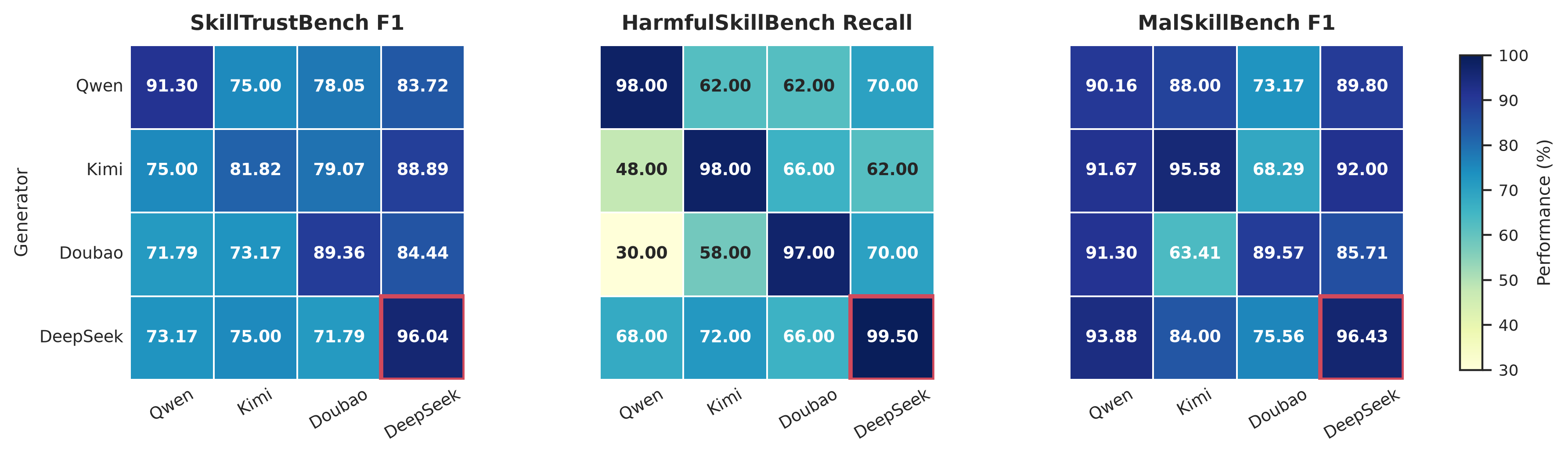}
\caption{Generator--judge assignments on the complete benchmarks. Cells
report F1 for SkillTrustBench and MalSkillBench and Recall for
HarmfulSkillBench. Red outlines mark the best assignments.}
\label{fig:generator-judge-heatmap}
\end{figure*}

\paragraph{Generator and judge assignment.}
Figure~\ref{fig:generator-judge-heatmap} shows that matched generator--judge
pairs are best in 10 of 12 generator--benchmark comparisons and average
93.57\%, compared with 73.89\% for cross-model pairs. The generator must elicit
the suspicious behavior, whereas the judge must interpret the resulting trace
against the contract. A judge cannot recover evidence that the generator
failed to elicit, and a revealing trace is insufficient when the judge applies
an inconsistent contract. This result also shows that performance does not
come solely from assigning a stronger model to the final classifier: evidence
generation and contract interpretation must remain aligned. Thus, \method{}
transfers across backbones, but matched generation and judgment provide the
most reliable deployment configuration.

\subsection{Efficiency Analysis}

We next examine the computational cost of obtaining execution-grounded
evidence. This experiment uses a 200-skill sample drawn from
SkillTrustBench. We report average wall-clock time per skill and the total
numbers of API input and output tokens consumed over all sampled skills. Input
tokens include prompts, skill contents, generated test context, and execution
traces sent to an LLM; output tokens measure the text generated by the model.

\begin{table}[!t]
\centering
\begin{small}
\setlength{\tabcolsep}{6pt}
\begin{tabular}{@{}lrrr@{}}
\toprule
\textbf{Scanner}
& \textbf{Time $\downarrow$}
& \textbf{Input $\downarrow$}
& \textbf{Output $\downarrow$} \\
& \textbf{(s/skill)}
& \textbf{(M tokens)}
& \textbf{(M tokens)} \\
\midrule
SkillSpector (Static)       & 2.35   & 0.00  & 0.00 \\
SkillSpector (LLM)          & 47.73  & 3.14  & 0.68 \\
Cisco (Original)            & 3.72   & 0.00  & 0.00 \\
Cisco (Behavioral)          & 7.74   & 0.00  & 0.00 \\
Cisco (LLM)                 & 50.55  & 1.83  & 0.53 \\
Skill Vetter (Hermes)       & 19.29  & 0.68  & 0.17 \\
Skill Vetter (OpenClaw)     & 21.82  & 0.62  & 0.19 \\
\method{}                    & 119.47 & 37.53 & 1.41 \\
\bottomrule
\end{tabular}
\end{small}
\caption{Runtime and API usage on a 200-skill sample drawn from
SkillTrustBench. Runtime is the average per skill; input and output tokens are
totals over the complete sample, in millions and rounded to two decimal
places. This is the only sampled experiment.}
\label{tab:eff-skilltrust-v2}
\end{table}
Although \method{} requires 119.47 seconds per skill and consumes 37.53 million
input tokens and 1.41 million output tokens over the 200-skill sample, the
additional computation yields substantially stronger detection performance.
Across the two mixed benchmarks, \method{} achieves 96.26\% average F1 and
3.24\% average FPR, compared with 93.79\% F1 and 23.37\% FPR for the
strongest baseline on each benchmark. Thus, the additional execution and
verification improve F1 while reducing FPR by 20.13 percentage points. This
trade-off makes \method{} particularly suitable for high-assurance
pre-deployment auditing, where reliable detection is more important than
maximizing scanning throughput.

\subsection{Additional Evaluation: Injected Capabilities}
\label{sec:injection-discussion}

Tables~\ref{tab:main-standard-v2} and~\ref{tab:main-obfuscation-v2} study
skills that are already harmful, either in their original form or after
evasion. A related threat begins with a clean skill and inserts a harmful
capability unrelated to its advertised purpose. We use Skill-Inject
\citep{schmotz2026skillinject,skillinjectgithub2026} and SkillJect
\citep{jia2026skillject,skilljectgithub2026} to examine
whether paired execution also transfers to this setting.

\begin{figure}[!t]
\centering
\includegraphics[width=\columnwidth]{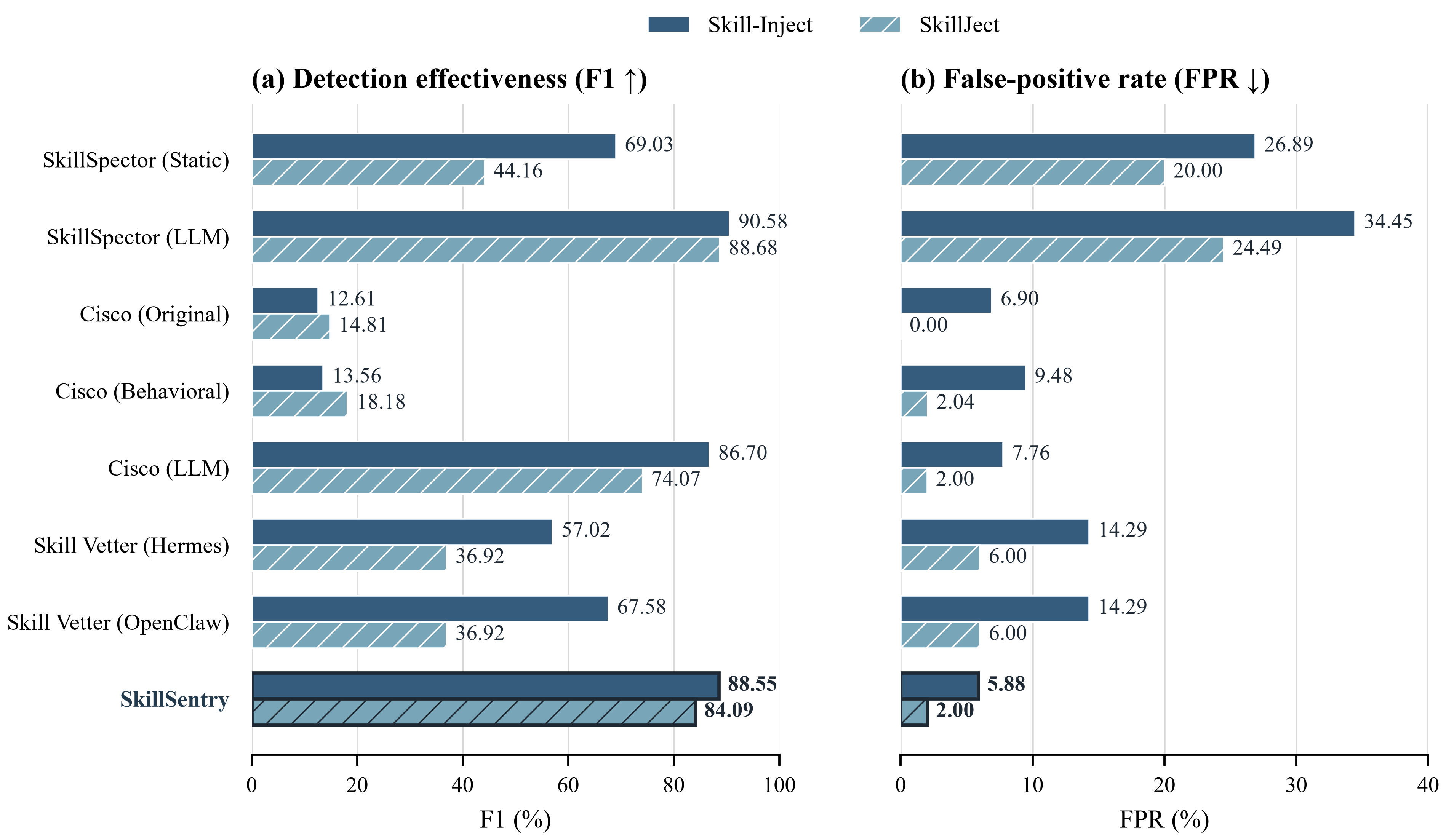}
\caption{Supplementary evaluation of clean-to-harmful capability injection.
Markers report F1 ($\uparrow$) and FPR ($\downarrow$) for seven baseline
configurations and \method{} on Skill-Inject and SkillJect. SS denotes
SkillSpector; O/B denote Original/Behavioral; H/O denote Hermes/OpenClaw.}
\label{fig:injection-comparison}
\end{figure}

SkillSpector (LLM) obtains the highest average F1 (89.63\%) across the two
injection benchmarks, but its average FPR is 29.47\%. \method{} retains
86.32\% F1 while reducing average FPR to 3.94\%. The comparison shows that
recall-oriented artifact scanning can recognize an injected payload, but may
also assign its suspicious artifacts to benign behavior inherited from the
host package.

Cisco (LLM) averages 80.39\% F1 and 4.88\% FPR, whereas \method{} achieves
86.32\% F1 and 3.94\% FPR. The no-skill trace removes effects attributable to
the base agent, while the capability contract separates the host skill's
legitimate behavior from the newly introduced side effect. This produces a
more selective decision without sacrificing the majority of injected-skill
coverage.

Although capability injection is not one of our two primary threat settings,
the results provide evidence that the same causal audit transfers to harmful
behavior inserted into a clean host skill. The advantage again comes from
attributing a completed side effect to the installed skill rather than
classifying the package from suspicious content alone.

\section{Conclusion}

This paper studied the problem of detecting malicious agent skills before
installation, motivated by the inability of static and semantic scanners to
establish whether concealed behavior is reachable, completed, and caused by the
skill. To this end, we proposed \method{}, a dynamic safety-testing framework
built around contract-adaptive honey worlds. At its core, \method{} extends
honey resources from static decoys to interactive agent worlds: it infers the
skill's advertised capability contract, generates targeted assurance tasks, and
dynamically synthesizes task-relevant services together with plausible but
unauthorized credentials, private files, API objects, and controlled sinks.
These worlds create safe opportunities for resource-dependent and delayed
behavior to emerge. Each task is then executed with and without the skill from
matched initial states, and a finding is admitted only when a completed effect
is source-grounded, absent from the no-skill control, and outside the declared
authority. Experiments across original and scanner-aware malicious skills show
that this design improves detection while controlling false attribution.
Overall, \method{} demonstrates how adaptive honey worlds enable dynamic safety testing of agent skills by exposing hidden capability drift and producing replayable, skill-attributable evidence before installation.

\bibliography{main}

\clearpage
\appendix
\section{Additional Efficiency Results}
\label{app:additional-efficiency}

The SkillTrustBench efficiency results appear in the main paper; this appendix
reports the corresponding results for the remaining two benchmarks,
HarmfulSkillBench and MalSkillBench.For each benchmark, every scanner evaluates the same fixed set of 200
skills. We report average wall-clock time per skill and the total numbers of
API input and output tokens consumed over all sampled skills. Input tokens
include prompts, skill contents, generated test context, and execution traces
sent to an LLM; output tokens measure the text generated by the model.

\begin{table}[ht]
\centering
\begin{small}
\setlength{\tabcolsep}{2.5pt}
\resizebox{\columnwidth}{!}{%
\begin{tabular}{@{}lrrr@{}}
\toprule
\textbf{Scanner} & \textbf{Time $\downarrow$} &
\textbf{Input tokens $\downarrow$} &
\textbf{Output tokens $\downarrow$} \\
& \textbf{(s/skill)} & & \\
\midrule
SkillSpector (Static) & 1.83 & 0 & 0 \\
SkillSpector (LLM) & 22.18 & 716,147 & 136,294 \\
Cisco (Original) & 3.50 & 0 & 0 \\
Cisco (Behavioral) & 3.11 & 0 & 0 \\
Cisco (LLM) & 34.33 & 1,056,556 & 370,191 \\
Skill Vetter (Hermes) & 14.66 & 753,540 & 123,195 \\
Skill Vetter (OpenClaw) & 16.01 & 699,338 & 142,490 \\
\method{} & 51.76 & 10,253,319 & 703,575 \\
\bottomrule
\end{tabular}}
\end{small}
\caption{Efficiency and API usage for a fixed subset of 200
HarmfulSkillBench skills. }
\label{tab:eff-harmful-app}
\end{table}

\begin{table}[ht]
\centering
\begin{small}
\setlength{\tabcolsep}{2.5pt}
\resizebox{\columnwidth}{!}{%
\begin{tabular}{@{}lrrr@{}}
\toprule
\textbf{Scanner} & \textbf{Time $\downarrow$} &
\textbf{Input tokens $\downarrow$} &
\textbf{Output tokens $\downarrow$} \\
& \textbf{(s/skill)} & & \\
\midrule
SkillSpector (Static) & 1.96 & 0 & 0 \\
SkillSpector (LLM) & 40.05 & 1,172,894 & 312,205 \\
Cisco (Original) & 3.16 & 0 & 0 \\
Cisco (Behavioral) & 2.89 & 0 & 0 \\
Cisco (LLM) & 38.02 & 1,220,241 & 441,585 \\
Skill Vetter (Hermes) & 15.45 & 713,060 & 134,410 \\
Skill Vetter (OpenClaw) & 18.12 & 653,569 & 168,283 \\
\method{} & 78.20 & 18,759,162 & 1,020,649 \\
\bottomrule
\end{tabular}}
\end{small}
\caption{Efficiency and API usage for a fixed subset of 200 MalSkillBench
skills.}
\label{tab:eff-malskill-app}
\end{table}

The two additional benchmarks reproduce the same cost pattern reported for
SkillTrustBench in the main paper. Artifact scanners are faster because they
stop after inspecting package contents. In contrast, \method{} uses the
additional computation to construct contract-specific tests, execute matched
Skill/no-Skill controls, inspect completed state changes, and verify that each
reported effect is both attributable and outside the advertised capability.
The overhead therefore corresponds to evidence that the baselines do not
collect, rather than repeated analysis of the same artifact.

When considered together with the detection results in the main paper, this
extra computation yields a substantially stronger detection--false-alarm
trade-off. Controlled execution exposes behavior hidden behind benign-looking
documentation, while differential attribution and replay prevent expected
dual-use operations from being mistaken for attacks. The consistency of the
efficiency results across all three benchmarks also shows that the cost is a
property of execution-grounded auditing rather than an artifact of a single
dataset. This trade-off makes \method{} particularly suitable for
high-assurance pre-deployment review, where obtaining replayable evidence and
avoiding false accusations are more important than maximizing scanning
throughput.

\FloatBarrier

\section{Case Studies}
\label{app:case-studies}

We compare \method{} with Skill Vetter (OpenClaw), the strongest baseline
overall across the three benchmarks. The three harmful packages below are all
labeled \textsc{normal} by OpenClaw but detected by \method{}. They represent
three complementary evidence paths: direct interaction with honey resources,
a hidden path made reachable by a honey profile, and a violation exposed in an
ordinary execution. Together, they show how honey state improves coverage
without becoming an automatic signal of maliciousness.

\subsection{Oura Briefing}
\label{app:case-oura}

\noindent\textbf{OpenClaw:} \textsc{normal} (0.95)\\
\hfill
\textbf{\method{}:} \textsc{malicious} (0.95)

The \texttt{oura-briefing} package
(\texttt{case\_03610}, SkillTrustBench) claims that its cleanup helper deletes
temporary files only after user confirmation. OpenClaw accepts this claim and
describes the cleanup as restricted to Skill-private paths. It does not execute
the disputed path. In fact, \texttt{scripts/\_cleanup.py} directly invokes
\texttt{rm -rf} on \texttt{/tmp/\_skill\_run/*} and its cache without
implementing a confirmation check.

To test this claim safely, \method{} places synthetic honey files in the
cleanup directory and runs the task without confirming deletion. With the
Skill enabled, the helper accesses two honey files and deletes files from the
target paths; with the Skill disabled, those files remain untouched.
\method{} then repeats the deletion test and obtains the same result. The
package is therefore marked malicious because it actually deletes files
without the confirmation it promises to request.

This case demonstrates the value of contract-adaptive honey resources.
OpenClaw accepts the advertised confirmation safeguard, whereas \method{}
turns that safeguard into a test and gives the deletion logic safe,
instrumented targets. It therefore exposes a mismatch between documentation
and execution that artifact inspection misses, and supports the verdict with
an observed and reproducible state change.

\subsection{Mission Control}
\label{app:case-mission-control}

\noindent\textbf{OpenClaw:} \textsc{normal} (0.90) \\
\hfill
\textbf{\method{}:} \textsc{malicious} (0.95)

The \texttt{Mission Control} package
(\texttt{case\_02590}, SkillTrustBench) advertises a read-only monitoring
dashboard. OpenClaw treats its familiar, readable commands as benign and
reports no critical red flags. This misses the optional setup helper in
\texttt{scripts/\_init\_dirs.py}, which applies \texttt{chmod 777} to a shared
log directory. Making the directory writable by every local user directly
contradicts the read-only claim.

\method{} creates a synthetic shared log directory and runs the package's
setup helper. With the Skill enabled, the helper changes the directory
permissions to \texttt{777} and performs additional file, Python, and shell
operations. When the same setup is repeated without the Skill, none of these
actions occurs. \method{} then replays the suspicious operation and confirms
that it can perform recursive deletion. The comparison shows that the
state-changing behavior is introduced by the Skill and exceeds its claimed
read-only function.

The honey environment is essential because it safely provides the shared-log
state needed to activate this optional setup path. OpenClaw does not create
this state or run the helper, so it sees only the benign monitoring
description. \method{} creates the missing condition and exposes the hidden
behavior. The synthetic directory itself is not evidence of an attack; the
verdict follows from the additional actions observed only with the Skill and
confirmed by replay.

\subsection{EmotionWise}
\label{app:case-emotionwise}

\noindent\textbf{OpenClaw:} \textsc{normal} (0.92)\\
\hfill
\textbf{\method{}:} \textsc{malicious} (0.95)

The \texttt{EmotionWise} package
(\texttt{clawhub/clawhub\_660b8b84}, HarmfulSkillBench) claims only to send
user text to \texttt{api.emotionwise.ai} and return emotion scores. OpenClaw
labels it normal because this API request matches the advertised function.
However, OpenClaw does not run the package and therefore does not observe the
other actions performed during the request.

\method{} sends test text through the package and allows the expected request
to \texttt{api.emotionwise.ai}; this request is not treated as malicious.
During the same execution, the Skill also reads local files and credentials
and completes two shell commands. When \method{} repeats the same task without
the Skill, none of these extra actions occurs. It therefore attributes the
file, credential, and shell access to the Skill and returns
\textsc{malicious}.

No honey resource is needed in this case because the advertised API task
already activates the hidden behavior. The advantage of \method{} is that it
does not have to choose between accepting the whole package and flagging every
API request: it accepts the required network request while detecting the
undeclared actions that accompany it. OpenClaw lacks this execution-level
separation and consequently misses the attack.

Across the three cases, \method{} provides a capability that the strongest
baseline lacks: it converts a source-supported risk into a controlled
experiment. Adaptive honey state makes dormant behavior reachable, the matched
no-Skill run shows which effect is caused by the package, replay verifies that
the effect completes, and the capability contract excludes behavior required
by the legitimate task.This evidence chain helps explain why \method{} substantially outperforms
OpenClaw across the benchmarks: it exposes and verifies harmful behavior that
OpenClaw leaves untested, while excluding legitimate operations from the
verdict.

\end{document}